\documentclass[iop]{emulateapj}
\usepackage{color}
\usepackage{graphicx}
\usepackage{amsmath}

\shorttitle{Testing the speed of light}
\shortauthors{Cao et al.}

\begin{document}
\title{Testing the speed of light over cosmological distances: the combination of strongly lensed and unlensed supernova Ia }

\author{Shuo Cao\altaffilmark{1}, Jingzhao Qi\altaffilmark{1}, Marek Biesiada\altaffilmark{1,2$\ast$}, Xiaogang Zheng\altaffilmark{1}, Tengpeng Xu\altaffilmark{1},  and Zong-Hong Zhu\altaffilmark{1$\dag$}}

\altaffiltext{1}{Department of Astronomy, Beijing Normal University,
Beijing 100875, China; \emph{zhuzh@bnu.edu.cn}}
\altaffiltext{2}{Department of Astrophysics and Cosmology, Institute
of Physics, University of Silesia, 75 Pu{\l}ku Piechoty 1, 41-500
Chorz{\'o}w, Poland; \emph{marek.biesiada@us.edu.pl}}

\begin{abstract}

Probing the speed of light is as an important test of General
Relativity but the measurements of $c$ using objects in the distant
universe have been almost completely unexplored. In this letter, we
propose an idea to use the multiple measurements of galactic-scale
strong gravitational lensing systems with type Ia supernova acting
as background sources to estimate the speed of light. This provides
an original method to measure the speed of light using objects
located at different redshifts which emitted their light in a
distant past. Moreover, we predict that strongly lensed SNe Ia
observed by the LSST would produce robust constraints on $\Delta
c/c$ at the level of $10^{-3}$. We also discuss whether the future
surveys such as LSST may succeed in detecting any hypothetical
variation of $c$ predicted by theories in which fundamental
constants have dynamical nature.

\end{abstract}


\maketitle


\section{Introduction}

As one of the cornerstones of General Relativity as well as of many
other gravitational theories, the speed of light $c$ plays a very
important role in modern physics. On the other hand, during the last
two decades a great attention has been paid to the theories with
varying speed of light (VSL), in which the speed of light might be
dynamical and could have been varying in the past. Any possible
violation of constancy of $c$ would have far-reaching consequences
for our understanding of the Nature. The interest in VSL theories
was triggered by the conjecture that they could become an
alternative to the inflationary mechanism of solving standard
cosmological problems \citep{Albrecht99,Barrow99a,Barrow99b}. On the
tide of growing popularity of VSL approach, \citet{Davies02} claimed
that variation of the speed of light can be discriminated from
variation of the elementary charge based on the entropy of black
holes. Later this claim was refuted by \citet{Carlip02,Flambaum02}.
It is worth recalling that \citet{Ellis05} gave a sobering review of
serious conceptual problems one faces while trying to change the
status of c in Physics. In particular, they stressed that it is
usually not consistent to allow a constant to vary in an equation
that has been derived from a variational principle under the
hypothesis of this quantity being constant. Therefore, one needs to
go back to the Lagrangian and derive new equations after having
replaced the constant by a dynamical field.

Although $c$ has been measured with a very high precision most of
these measurements were carried out on Earth or in our close cosmic
surroundings. Recently, however, some advances have been made
concerning measurements of the speed of light using extragalactic
objects. Considering the relation between the maximum value of the
angular diameter distance $D_A(z)$ and the Hubble parameter $H(z)$
corresponding to this maximum, \citet{Salzano15} discussed (on the
simulated data) the possibility of using the Baryon Acoustic
Oscillations (BAO) to place constraints on the variation of the
speed of light. Then, \citet{Cai16} investigated the constancy of
$c$ using independent observations of Hubble parameters $H(z)$ and
luminosity distances from type Ia supernovae (SNe Ia)
\citep{Suzuki12}. The measurements of $c$ in the distant universe is
an almost completely uncharted territory. Recently, with the angular
diameter distances measured for intermediate luminosity quasars
extending to high redshifts, \citet{Cao17a} performed the first
measurement of the speed of light referring to the redshift baseline
$z=1.70$. The result was in a very well agreement with the value of
$c$ obtained on the Earth (i.e. at $z=0$).

Gravitational lensing creates another opportunity to test the speed
of light. In the well known gravitational lens Q0957+561
\citep{Walsh79,Young81}, the appearance of two images of $z_s=1.41$
quasar within an Einstein radius of $3''.08$ around the core of the
lensing galaxy at $z_l=0.36$, has opened up a wide range of
possibilities of using strong lensing systems in cosmology and
astrophysics
\citep{Grillo08,Biesiada10,Cao12a,Cao12b,Cao12c,Cao13,Cao14,Cao15,Cao16a,Li16,Cao17c}.
Strongly lensed SNe Ia, which has long been predicted in the
literature \citep{Refsdal64}, remained un-discovered until very
recently. Eventually, \citet{Goobar17} reported the discovery of a
new gravitationally lensed type Ia supernova (SNe Ia) iPTF16geu (SN
2016geu) from the intermediate Palomar Transient Factory (iPTF).
Strong lensing time-delay predictions for this system were discussed
in detail in \citet{More17}.

Here, we will focus on the idea of constraining $c$ by using the
measurements of galactic-scale strong gravitational lensing systems
with SNe Ia acting as background sources. In a specific strong
lensing system, time delay between the images of strongly lensed SN
Ia is directly related to the speed of light, the Fermat potential
difference, and the so-called time-delay distance. This opens a
possibility to measure the speed of light on the baseline up to the
redshift of the source.

\section{Methodology}

General Relativity predicts that light can be deflected by massive
objects (e.g. galaxies) resulting in distorted multiple images of
the background sources. For a specific strong-lensing system with
the lensing galaxy at redshift $z_l$, the multiple-image separation
of the source at redshift $z_s$ depends on the ratio of
angular-diameter distances between lens and source $D_{ls}$ and
between observer and source $D_{s}$ \citep{Schneider92}. In a
spatially flat Friedman-Lema\^{\i}tre-Robertson-Walker metric the
formula for the angular diameter distance reads
\begin{eqnarray}
\label{inted} D_A(z_1, z_2)=\frac{c_{z_s}}{H_{0}
(1+z_2)}\int_{z_1}^{z_2} \frac{dz'}{E(z')} \, ,
\end{eqnarray}
where $H_0$ is the Hubble constant and $E(z)$ is the dimensionless
expansion rate dependent on redshift $z$. It should be stressed at
this point that even though we mentioned VSL theories above, we do
not make use of any of these models. On the contrary, we use (here
and in the equations that follows) the standard theory perceiving
our goal as a measurement of the speed of light using extragalactic
objects. We use the expression $c_{z_s}$ instead of just $c$ in
order to emphasize that the speed of light would be determined on
the baseline between the observer $z=0$ and the source $z_s$. Hence,
if the measured value turns out $c_{z_s}=c$ within statistical and
systematic uncertainties, this would be a confirmation of the
constancy of $c$ (and hence the standard Physics that we know on
Earth) using very distant objects. On the contrary, if $c_{z_s} \neq
c$ significantly (note that this may happen in a space dependent
manner) this could be a signal that $c$ was not a fundamental
constant, triggering further theoretical work in order to explain
this result. At last, but not least, our approach makes it clear
that, although we report dimension-full $c_{z_s}$, we do really
constrain the dimensionless ratio $c/c_0$ or $\Delta c/c$.


Provided that the background source is a SNe Ia which is a transient
event with well defined light curve after peak, the strong lensing
time delay effect due to different light paths combined with Shapiro
effect will be revealed in the photometry of its multiple images.
Recently, strong gravitational time delays between the multiple
images have become a promising tool in cosmology providing
alternative measurements of the Hubble constant
\citep{HOLI,Liao2017}. According to the theory of gravitational
lensing, time delay between images $\boldsymbol{\theta}_i$ and
$\boldsymbol{\theta}_j$ can be written as \citep{Treu10}
\begin{equation}
\Delta t_{i,j} = \frac{D_{\mathrm{\Delta
t}}(1+z_{\mathrm{l}})}{c_{z_s}}\Delta \phi_{i,j}, \label{relation}
\end{equation}
where
$\Delta\phi_{i,j}=[(\boldsymbol{\theta}_i-\boldsymbol{\beta})^2/2-\psi(\boldsymbol{\theta}_i)-(\boldsymbol{\theta}_j-\boldsymbol{\beta})^2/2+\psi(\boldsymbol{\theta}_j)]$
is the Fermat potential difference determined by the lens mass
distribution. The source position, with respect to the line
connecting the observer and the center of the lens is denoted by
$\boldsymbol{\beta}$, and $\psi$ denotes two-dimensional lensing
potential fulfilling the Poisson equation $\nabla^2\psi=2\kappa$,
where $\kappa$ is dimensionless surface mass density (convergence)
of the lens. The time-delay distance is defined as
\begin{equation}
D_{\mathrm{\Delta t}}=\frac{D_{\mathrm{l}}
D_{\mathrm{s}}}{D_{\mathrm{ls}}}.
\end{equation}
where $D_{ls}$ and $D_s$ are angular diameter distances between the
lens and the source and between the observer and the source.

Assuming flat Friedman-Robertson-Walker metric, one can relate the
angular diameter distance $D_A$ to the proper distance $D_P$
according to $D_A(z_1,z_2)=D_P(z_1,z_2)/(1+z_{2})$. Since proper
distances are additive, the distance ratio of $D_{ls}/D_s$ can be
expressed in terms of angular diameter distances $D_{l}$ and $D_s$
\begin{equation}\label{ratio}
 \frac{D_{ls}}{D_s}=1-\frac{1+z_l}{1+z_s}\frac{D_l}{D_s}
\end{equation}
By combining Eq.~(1)-(\ref{ratio}), one can express the speed of
light as
\begin{equation}
c_{z_s} =
\frac{D_l(1+z_l)D_s(1+z_s)}{(1+z_s)D_s-(1+z_l)D_l}\frac{1}{\Delta
t_{i,j}}\Delta \phi_{i,j}, \label{cmeasurement}
\end{equation}
which suggests how to measure this quantity using strong lensing
systems. Note that for each specific strong lensing system, the
derivation of Eq.~(2)-(5) is based on the choice that $c_{z_s}$
denotes the constant speed of light related to the baseline from the
source to the observer. It is also straightforward to check that
when the speed of light is a function of redshifts, the accurate
derivation of the above equations could be very different, since the
cosmological distances contain more a factor $c(z)$ outside the
integral \citep{Qi14}.

Current observational techniques allow the redshifts of the lens
$z_{\mathrm{l}}$ and the source $z_{\mathrm{s}}$ to be measured
precisely. Moreover, imaging and spectroscopy from the Hubble Space
Telescope (HST) and ground-based observatories make it possible to
derive three key ingredients for individual lenses: stellar velocity
dispersion, high-resolution images of the lensing systems, and time
delays. On the one hand, current high-resolution image astrometry,
combined with the state-of-the-art lens modelling techniques
\citep{Suyu10,Suyu12b} and kinematic modelling methods
\citep{Auger10,Sonnenfeld12}, will precisely determine the multiple
image positions of the background source, the Einstein radius of the
lensing system and thus place stringent limits on the Fermat
potential $\Delta \phi_{i,j}$ with $<3\%$ uncertainty (including the
systematics) \citep{Suyu13,Suyu14,Liao2015}. In practice, we also
need to quantify the influence of matter along the line of sight
(LOS) on the lens potential, contributing an extra systematic
uncertainty at $1\%$ level~\citep{HOLI}. On the other hand, in the
framework of typical quasar-elliptical galaxy lensing systems,
well-measured light curves combined with new curve shifting
algorithms \citep{Tewes13a} could provide $\Delta t_{i,j}$
measurements typically with 3\% accuracy
\citep{Fassnacht02,Courbin11,Tewes13b,Liao2015,Greg2015}. Time
delays measured in lensed SNe Ia are supposed to be very accurate
due to exceptionally well-characterized spectral sequences and
considerable variation in light curve morphology
\citep{Nugent02,Pereira13}. In this analysis, in the framework of
 SNe Ia-elliptical galaxy lensing systems, the fractional
uncertainty of $\Delta t$ is taken at the level of 1\%, which is a
reasonable assumption for well-measured light curves of lensed SNe
Ia.

Concerning the time-delay distance, one might be tempted to
pre-assume cosmological model e.g. as flat $\Lambda$CDM with
$\Omega_m = 1- \Omega_{\Lambda} =0.3$ but this would introduce a
hardly controllable bias. More reasonable approach is to derive the
distances to the lens and to the source based on their redshifts and
using the absolute distances to standard candles at these redshifts.
It is commonly believed that unlensed SNe Ia can be calibrated as
standard candles, which, combined with the so-called
distance-duality relation \citep{Etherington33,Cao11,Cao16b}, can
provide the $D_A(z)$ (in the unit of Mpc) both at lens and source
redshifts:
\begin{equation}
D_A(z)=\left(\frac{1}{1+z}\right)^2 10^{(m_{X}-M_B-K_{BX})/5-5}.
\end{equation}
where $m_X$ is the peak apparent magnitude of the supernova in
filter $X$, $M_B$ is its rest-frame B-band absolute magnitude, and
$K_{BX}$ denotes the cross-filter K-correction \citep{Kim96}. For
the purpose of our analysis, we determined the angular diameter
distances $D_l$ and $D_s$ of strongly lensed SNs Ia by fitting a
polynomial to the unlensed SNe Ia data. Therefore we bypassed the
need to pre-assume any specific cosmological model. We expect that,
in the near future, much bigger depth, area, resolution and sample
sizes brought by the next generation wide and deep sky surveys
\citep{Marshall05} will yield a huge number of strong lensing
systems discovered, including lensed SNe Ia. In the following, we
will illustrate what kind of results one could get using the future
data from the forthcoming Large Synoptic Survey Telescope (LSST).

\begin{figure}
\includegraphics[scale=0.4]{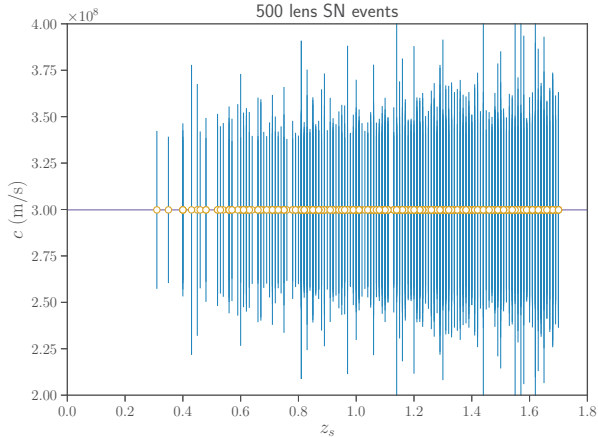}
\caption{ Individual measurements of the speed of light from the
forthcoming LSST survey.}\label{fig1}
\end{figure}

\begin{figure}
\includegraphics[scale=0.4]{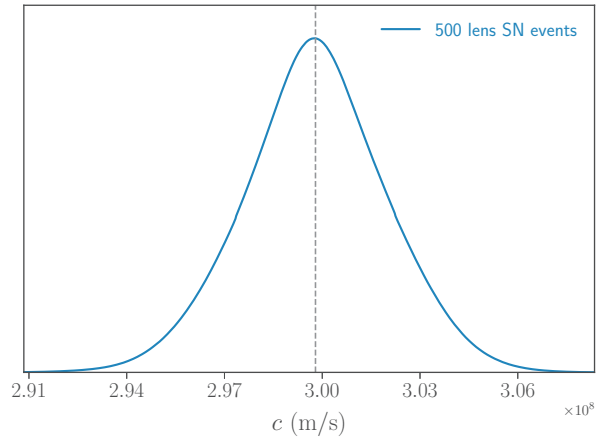}
\caption{ Probability distribution of the speed of light $c$
possible to obtain from the forthcoming LSST survey. }\label{fig2}
\end{figure}

\begin{table*}
\begin{center}
\begin{tabular}{lcccc}
\hline\hline

& $\delta\Delta t$ & $\delta\Delta\psi$ &  $\delta LOS$  &\\
\hline
Lensed SNe Ia  &1\% & 3\% &1\% & \\
\hline & $\sigma_{\rm meas}~\rm{(mag)}$  & $\sigma_{\rm int}~\rm{(mag)}$ & $\sigma_{\rm lens}~\rm{(mag)}$ & $\sigma_{sys}~\rm{(mag)}$ \\
\hline
Unlensed SNe Ia & 0.08  & 0.09  & 0.07 $\times z$ & $0.01(1 + z)/1.8$\\
\hline\hline
\end{tabular}
\end{center}
\caption{First row shows relative uncertainties of factors
contributing to the accuracy of $c$ measurement. $\delta\Delta t$,
$\delta\Delta\psi$, $\delta LOS$ correspond to time delay, Fermat
potential difference and light-of-sight contamination, respectively.
Second row displays different factors contributing to the
uncertainty of the magnitude of unlensed SNe Ia.  $\sigma_{\rm
meas}$, $\sigma_{\rm int}$, $\sigma_{\rm lens}$, $\sigma_{sys}$
denote the photometric measurement uncertainty, the intrinsic
scatter, the lensing magnification uncertainty and the total
systematic uncertainty.}\label{error}
\end{table*}

\section{Simulated data}

Within the next decade new wide-area imaging surveys are likely to
discover thousands of lensed SNe Ia, time delay measurements of
which will be derived from their time-domain information (observed
by repeated scan of sky) and dedicated follow-up monitoring
campaigns. Detailed calculation of the likely yields of several
planned strong lensing surveys was recently performed by
\citet{Goldstein17}. Considering realistic simulation of lenses and
sources, they found that LSST can discover up to 500 multiply imaged
SNe Ia in a 10 year $z$-band search, which is more than an order of
magnitude improvement over previous estimates \citep{Oguri10}, while
the LSST is expected to yield $10^6$ unlensed type-Ia supernovae
\citep{Cullan17}.

In the framework of the method proposed by \citet{Collett15}, we
first simulated a population of realistic strong lenses possible to
be observed by LSST. Because elliptical galaxies dominate in the
galaxy lensing cross section \citep{Oguri10}, we considered only SN
Ia lensed by early-type galaxies, whole velocity dispersion function
in the local Universe follows the modified Schechter function
\citep{Choi07} (for discussion about such choice in view of other
data on velocity dispersion distribution functions see
\citet{Cao12a,Biesiada14}). In order to assess the accuracy of the
Fermat potential recovery, we assumed the elliptically-symmetric
power-law lens model, which has been widely used in several studies
of X-ray observations of galaxies \citep{Humphrey10} as well as
strong lensing caused by early-type galaxies
\citep{Koopmans06,Treu06,Gavazzi07,Koopmans09}. Following previous
works \citep{Oguri08,Oguri10}, we assumed a Gaussian distribution
for the ellipticity $e \sim {\cal N}(0.3,0.16)$. Meanwhile, lens
velocity dispersions are directly related to their masses. So, in
order to check how well does our simulation represent real lenses we
found that our simulated population of lenses is dominated by
galaxies with $\sigma_{v}\sim 200$ km/s, having approximately
Gaussian distribution characterized by $\sigma_{v}=210\pm50$ km/s.
These parameters could be inferred from high-quality imaging
observations through state-of-the-art lens modelling techniques
\citep{Suyu10,Suyu12b} and kinematic modelling methods
\citep{Auger10,Sonnenfeld12}. Comparison of this result with the
SL2S sample, concerning the distribution of the velocity dispersion
in the population of lenses, reveals similarity between the
simulations and the real observations \citep{Sonnenfeld13a}.

In order to produce a catalog of lensed SN Ia candidates, three
sources of uncertainties are included in our simulation of lensed
SNe Ia: time delay, Fermat potential difference and the line of
sight (LOS) effect \citep{Liao2017}. Firstly, compared with strongly
lensed galaxies and quasars, which have already been discovered in
large number \citep{Cao15,Shu17}, strongly lensed SNe Ia have
notable advantages over traditional strong lenses as time-delay
indicators. More specifically, benefiting from exceptionally
well-characterized spectral sequences and relatively small variation
in quickly evolving light curve shapes and color
\citep{Nugent02,Pereira13}, time delays measured from lensed SNe Ia
are less onerous than from AGNs and quasars. Therefore, as pointed
out in the recent analysis by the strong lens time delay challenge
(TDC) \citep{Liao2015,Greg2015}, the fractional uncertainty at the
level of 1\% can be achieved for $\Delta t$. Secondly, a
high-resolution imaging while the SNe Ia is still active is
necessary to precisely determine the lens potential. Our simulation
of a system with the lensed SNe Ia image quality typical to the HST
observations and recovery of the relevant parameters with
state-of-the-art lens modelling techniques, demonstrates that one
can achieve the lens modelling (i.e., the Fermat potential
difference) precision at $\sim 3\%$ level for a single well-measured
time-delay lens system \citep{HOLI}. Finally, despite of these
advantages, lensed SNe Ia still face the problem of the line of
sight (LOS) contamination. Using the technique of simulations of
many multiple image configurations, one could be able to quantify
the influence of the matter along the LOS on strong lensing systems
\citep{Jaroszyki12}. Further progress in this direction has been
achieved by \citet{panglos}, with the publicly available code {\tt
Pangloss} reconstructing the mass along a line of sight up to
intermediate redshifts. We assumed that the LOS contamination might
introduce 1\% uncertainty in the lens potential \citep{Liao2017}.

We performed a Monte Carlo simulation to create the unlensed SNe Ia
sample. The simulation was carried out in the following way: I) When
calculating the sampling distribution (number density) of the SNe Ia
population, we adopted the redshift-dependent SNe Ia rate from
\citet{Sullivan00}. In each simulation, there were 5000 type Ia
supernova covering the redshift range of $0.00<z \leq 1.70$. II)
Following the suggestion of \citet{Goldstein17}, the peak
rest-frame $M_B$ was assumed normally distributed with the mean -19.3 and 
standard deviation 0.2, while the cross-filter $K$-corrections were
computed from the one-component SNe Ia spectral template of
\citet{Nugent02} (see \citet{Barbary14} for more details). III)
Three sources of uncertainties were included in the peak apparent
magnitude of the supernova. First, following the strategy described
by the WFIRST Science Definition Team (SDT) \citep{Spergel15}, the
distance precision per SNe was determined by the following
uncertainty model: $\sigma^2_{\rm stat}= \sigma^{2}_{\rm meas}+
\sigma^{2}_{\rm int} + \sigma^{2}_{\rm lens}$ \citep{Hounsell17},
with the mean uncertainty (including both statistical measurement
uncertainty and statistical model uncertainty) $\sigma_{\rm
meas}=0.08$ mag, the intrinsic scatter uncertainty $\sigma_{\rm int}
= 0.09$~mag, and the lensing uncertainty set to be $\sigma_{\rm
lens} = 0.07 \times z$~mag \citep{Holz05,Jonsson10}. Moreover, the
total systematic uncertainty was also considered in the corrected
SN~Ia distances, which is modeled as $\sigma_{sys} = 0.01(1 +
z)/1.8~\rm{(mag)}$ \citep{Hounsell17}.

In Table I we list the relative or absolute uncertainties of the
above mentioned factors contributing to the accuracy of $c$
measurement. Assuming that parameters whose uncertainties listed in
Table I follow Gaussian distributions, one is able to obtain $c$ and
its average value using Eq.~(5). In order to guarantee unbiased
final results, this process was repeated 10$^3$ times for each
strong lensing system.

\section{Results and discussion}

Expected results of the speed of light measurements obtained from
LSST are shown in Fig.~1.
Previous successful measurement of $c$ using the quasar sample
combined with the expansion rate function $H(z)$ based on the data
from cosmic chronometers and BAO (see \citet{Qi18} for the recent
observations of Hubble parameters) , even though being an important
achievement providing the first assessment of the speed of light
referring to the distant past, was just a single measurement with
the base-line $z=1.70$ \citep{Cao17a}. In contrast, concerning the
method proposed here, we expect that the high resolution imaging of
LSST will make it possible to discover a large number of strong
lensing systems leading to $N\sim500$ measurements of $c$ for the
optimal imaging case. The question now arises: \emph{Are these
measurements sufficient enough to detect possible VSL effects?}
Considering the high precision measurement of the speed of light
$c_0$ on Earth, with the fractional uncertainty of $10^{-9}$, it is
very difficult to achieve competitive results with cosmological
measurements. SNe Ia observations, however, would provide us the
value of $c_{z_s}$, the speed of light at redshift base-line $z_s$,
from which we may study the accuracy concerning the deviation from
$c_0$, $\Delta c=c_{z_s}-c_0$. The effectiveness of our method could
be seen from the precision of measurements of $c$ at different
redshift base-lines. These are summarized in Fig.~2 and Fig.~3. The
forecasts for the LSST survey is: strongly lensed SNe Ia observed by
LSST would produce robust constraints on the speed of light at the
level of $\Delta c/c=0.005$, if the distance measurements from
unlensed counterparts are available. This is the most unambiguous
result of the current dataset.

It is interesting to compare our results with the analysis of
\citet{Salzano15} who first proposed to test the constancy of $c$
using cosmological observables. They noticed that at the redshift
$z_M$ where the angular diameter distance attains its maximum
$D_A(z_M) H(z_M) = c_{z_M}$. Therefore, whenever one was able to
observationally determine $z_M$ and reconstruct (with some
non-parametric method) $D_A(z)$ in its neighborhood together with
the expansion rate $H(z)$, the measurement of $c_{z_M}$ could be
directly obtained. Then, their idea was to use the BAO measurements
providing both radial and tangential BAO modes to accomplish this
goal. More specifically, \citet{Salzano15} found that the future
missions such as Euclid would be able to give stringent fits on the
speed of light at the level of $\Delta c/c=0.009$, while a VSL of
1\% (if any) will be detect at 1$\sigma$ level. Further progress in
this direction has recently been achieved by \citet{Cai16}, who
showed that DES can not provide better improvement in detecting
variation of the speed of light for the 1\% case. Another approach
was taken in the paper of \citet{Cao17a}, where $D_A(z)$ function
was reconstructed (using Gaussian processes) from compact radio
sources and $H(z)$ from passively evolving galaxies. The result --
first actual measurement of the speed of light using extragalactic
objects -- confirmed constancy of $c$ at $6\%$ level (1$\sigma$
uncertainty level, or $1.4\%$ based on the central fit). Therefore,
the combination of strongly lensed and unlensed SN Ia may achieve
considerably higher precision of $c$ measurements than the other
popular astrophysical probes including BAO+$H(z)$. We would like to
stress that our method of measuring $c$ from lensed SNe Ia is not
only competitive with BAO, but also does not rely on any pre-assumed
values of cosmological parameters.

\begin{figure}
\includegraphics[scale=0.4]{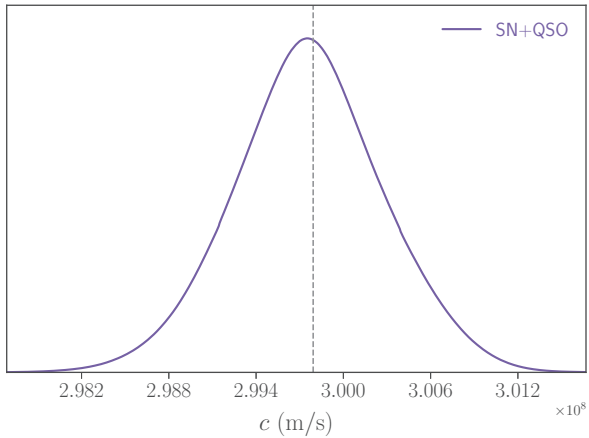}
\caption{ Probability distribution of the speed of light $c$
possible to obtain from the forthcoming LSST survey, with the
combination of lensed SNe Ia and lensed quasars. }\label{fig3}
\end{figure}

There are still many ways in which our technique might be improved.
First is suggested by the methodology applied in our previous study
\citep{Cao17a}. Hence, our method could also be applied to the
quasar-galaxy strong lensing systems with radio quasars acting as
background sources. According to the analysis of \citet{Oguri10},
LSST should find 3000 lensed quasars, an order of magnitude
improvement over the lensed SNe Ia. Instead of SNe Ia as standard
candles, one may use the angular sizes of the compact structure in
intermediate luminosity radio quasars \citep{Cao17a} from the
very-long-baseline interferometry (VLBI) observations and use them
as standard rulers at different redshifts \citep{Cao17b}. Because
statistical uncertainty is inversely proportional to $\sqrt{N}$,
where $N$ is the number of systems, future observations of a lensed
quasars will enable us to get more precise measurements of the speed
of light. Actually, such combination of strongly lensed SNe Ia and
quasars will results in more stringent constraints on the speed of
light at the level of $\Delta c/c=0.001$. The results are shown in
Fig.~3. Even more important is that in such case LSST may succeed in
detecting possible existence of VSL effects. Finally, let us
emphasize that the improved accuracy of the lens model and the
time-delay measurements is crucial to our method. Recently,
\citet{Liao2017} proposed a new strategy to simultaneously detect
strongly lensed gravitational waves (GWs) and their electromagnetic
(EM) counterpart, which will improve the precision of the Fermat
potential reconstruction to 0.5\%. Given the wealth of available
imaging and spectroscopic data, one might be optimistic about
achieving much higher precision of lensing time-delays in the
future. In that case, the precision of $c$ measurements using
extragalactic objects would be much higher.

\section*{Acknowledgments}

This work was supported by National Key R\&D Program of China No.
2017YFA0402600; the National Basic Science Program (Project 973) of
China under (Grant No. 2014CB845800); the National Natural Science
Foundation of China under Grants Nos. 11503001, 11690023, 11373014,
and 11633001; Beijing Talents Fund of Organization Department of
Beijing Municipal Committee of the CPC; the Strategic Priority
Research Program of the Chinese Academy of Sciences, Grant No.
XDB23000000; the Interdiscipline Research Funds of Beijing Normal
University; and the Opening Project of Key Laboratory of
Computational Astrophysics, National Astronomical Observatories,
Chinese Academy of Sciences. J.-Z. Qi was supported by China
Postdoctoral Science Foundation under Grant No. 2017M620661.
M.B. was supported by Foreign Talent Introducing Project and Special Fund
Support of Foreign Knowledge Introducing Project in China.



\begin{thebibliography}{}

\bibitem[Albrecht \& Magueijo(1999)]{Albrecht99} Albrecht, A. \& Magueijo, J. 1999, PRD, 59, 043516
\bibitem[Auger et al.(2010)]{Auger10} Auger, M. W., et al. 2010, ApJ, 724, 511
\bibitem[Barbary(2014)]{Barbary14} Barbary, K. 2014, doi:10.5281/zenodo.11938
\bibitem[Barrow(1999)]{Barrow99a} Barrow, J. D. 1999, PRD, 59, 043515
\bibitem[Barrow \& Magueijo(1999)]{Barrow99b} Barrow, J. D. \& Magueijo, J. 1999, PLB, 447, 246
\bibitem[Biesiada, Pi\'{o}rkowska, \& Malec(2010)]{Biesiada10} Biesiada, M., Pi\'{o}rkowska, A., \& Malec, B. 2010, MNRAS, 406, 1055
\bibitem[Biesiada, et al.(2014)]{Biesiada14} Biesiada, M., et al. 2014, JCAP, 10, 080.
\bibitem[Cai et al.(2016)]{Cai16} Cai, R. G., Guo, Z. K., \& Yang, T. 2016, JCAP, 08, 016
\bibitem[Cao \& Liang(2011)]{Cao11} Cao, S., \& Liang, N. 2011, RAA, 11, 1199
\bibitem[Cao \& Zhu(2012)]{Cao12a} Cao, S., \& Zhu, Z.-H. 20112, A\&A, 538, A43
\bibitem[Cao, Covone \& Zhu(2012)]{Cao12b} Cao, S., Covone, G., \& Zhu, Z.-H. 2012, ApJ, 755, 516
\bibitem[Cao et al.(2012)]{Cao12c} Cao, S., Pan, Y., Biesiada, M., God{\l}owski, W. \& Zhu, Z.-H. 2012, JCAP, 03, 016
\bibitem[Cao et al.(2013)]{Cao13} Cao, S., et al. 2013, RAA, 13, 15
\bibitem[Cao \& Zhu(2014)]{Cao14} Cao, S., \& Zhu, Z.-H. 2014, PRD, 90, 083006
\bibitem[Cao et al.(2015)]{Cao15} Cao, S., Biesiada, M., Gavazzi, R., Pi{\'o}rkowska, A. \& Zhu Z.-H. 2015, ApJ, 806, 185
\bibitem[Cao et al.(2016a)]{Cao16a} Cao, S., et al. 2016a, MNRAS, 461, 2192
\bibitem[Cao et al.(2016b)]{Cao16b} Cao, S., et al. 2016b, MNRAS, 457, 281
\bibitem[Cao et al.(2017a)]{Cao17c} Cao, S., et al. 2017a, ApJ, 835, 92
\bibitem[Cao et al.(2017b)]{Cao17a} Cao, S., Biesiada, M., Jackson, J., Zheng, X. \& Zhu Z.-H. 2017b, JCAP, 02, 012
\bibitem[Cao et al.(2017c)]{Cao17b} Cao, S., Zheng X., Biesiada M., Qi J., Chen Y. \& Zhu Z.-H. 2017c, A\&A, 606, A15
\bibitem[Carlip \& Vaidya(2002)]{Carlip02} Carlip, S., \& Vaidya, S. 2002, hep-th/0209249
\bibitem[Choi et al.(2007)]{Choi07}  Choi, Y.-Y., et al. 2007, ApJ, 658, 884
\bibitem[Collett et al.(2014)]{panglos} Collett, T. E., et al. 2013, MNRAS, 432, 679
\bibitem[Collett(2015)]{Collett15} Collett, T. E. 2015, ApJ, 811, 20
\bibitem[Cullan et al.(2017)]{Cullan17} Cullan, H., Robotham, A. S. G., Lagos, C. D. P., \& Kim, A. G., 2017, ApJ, 847, 128
\bibitem[Courbin et al.(2011)]{Courbin11} Courbin, F., et al. 2011, A\&A, 536, A53
\bibitem[Davies et al.(2002)]{Davies02} Davies, P. C. W., et al. 2002, Nature, 418, 602
\bibitem[Dobler et al.(2015)]{Greg2015} Dobler, G., et al. 2015, ApJ, 799, 168
\bibitem[Ellis \& Uzan(2005)]{Ellis05} Ellis, G. F. R., \& Uzan, J.-P. 2005, Am. J. Phys. 73, 240
\bibitem[Etherington(1933)]{Etherington33} Etherington, I. M. H. 1933, Philosophical Magazine, 15, 761 (reprinted in 2007, Gen. Rel. Grav. 39, 1055)
\bibitem[Fassnacht et al.(2002)]{Fassnacht02} Fassnacht, C. D., et al. 2002, ApJ, 581, 823
\bibitem[Flambaum(2002)]{Flambaum02} Flambaum, V. V. 2002, astro-ph/0208384
\bibitem[Gavazzi et al.(2007)]{Gavazzi07} Gavazzi, R., Treu, T., Rhodes, J. D., et al. 2007, ApJ, 667, 176
\bibitem[Goldstein \& Nugent(2017)]{Goldstein17} Goldstein, D. A., \& Nugent, P. E. 2017, ApJL, 834, L5
\bibitem[Goobar et al.(2017)]{Goobar17} Goobar, A., et al. 2017, Science, 356, 291
\bibitem[Grillo et~al.(2008)]{Grillo08} Grillo, C., Lombardi, M., \& Bertin, G. 2008, A\&A, 477, 397
\bibitem[Holz \& Hughes(2005)]{Holz05} Holz, D. E., \& Hughes, S. A., 2005, ApJ, 629, 15
\bibitem[Hounsell et al.(2017)]{Hounsell17} Hounsell, R., et al. 2017, arXiv:1702.01747v1
\bibitem[Humphrey \& Buote(2010)]{Humphrey10} Humphrey, P. J., \& Buote, D. A. 2010, MNRAS, 403, 2143
\bibitem[Jaroszy\'{n}ski, et al.(2012)]{Jaroszyki12} Jaroszy\'{n}ski, M., \& Kostrzewa-Rutkowska, Z. 2012, MNRAS, 424, 325
\bibitem[J\"{o}nsson et al.(2010)]{Jonsson10} J\"{o}nsson, J., et al. 2010, MNRAS, 405, 535
\bibitem[Kim et al.(1996)]{Kim96} Kim, A., Goobar, A., \& Perlmutter, S. 1996, PASP, 108, 190
\bibitem[Koopmans et al.(2006)]{Koopmans06} Koopmans, L. V. E., et al. 2006, ApJ, 649, 599
\bibitem[Koopmans et al.(2009)]{Koopmans09} Koopmans, L. V. E., et al. 2009, ApJ, 703, L51
\bibitem[Li et al.(2016)]{Li16} Li, X. L., et al. 2016, RAA, 16, 84
\bibitem[Liao et al.(2015)]{Liao2015} Liao, K., et al. 2015, ApJ, 800, 11
\bibitem[Liao et al.(2017)]{Liao2017} Liao, K., Fan, X.-L., Ding, X., Biesiada, M. \& Zhu Z.-H., 2017, Nature Communications, 8, 1148
\bibitem[Marshall et al.(2005)]{Marshall05} Marshall, P., Blandford, R., \& Sako, M. 2005, NAR, 49, 387
\bibitem[More et al.(2017)]{More17} More, A., et al. 2017, ApJL, 835, L25
\bibitem[Nugent et al.(2002)]{Nugent02} Nugent, P., Kim, A., \& Perlmutter, S. 2002, PASP, 114, 803
\bibitem[Oguri et al.(2008)]{Oguri08} Oguri, M., et al., 2008, AJ, 135, 512
\bibitem[Oguri \& Marshall(2010)]{Oguri10} Oguri, M. \& Marshall, P. J. 2010, MNRAS, 405, 2579
\bibitem[Pereira et al.(2013)]{Pereira13} Pereira, R., Thomas, R. C., Aldering, G., et al. 2013, A\&A, 554, A27
\bibitem[Qi et al.(2014)]{Qi14} Qi, J. Z., et al. 2014, PRD, 90, 063526
\bibitem[Qi et al.(2018)]{Qi18} Qi, J. Z., et al. 2018, RAA, 18, 66
\bibitem[Refsdal(1964)]{Refsdal64} Refsdal, S. 1964, MNRAS, 128, 307
\bibitem[Salzano et al.(2015)]{Salzano15} Salzano, V., Dabrowski, M. \& Lazkoz, R. 2015, PRL, 114, 101304
\bibitem[Schneider et~al.(1992)]{Schneider92} Schneider, P., Ehlers, J., \& Falco, E.~E. 1992, Gravitational Lenses
\bibitem[Seikel et al.(2012)]{Seikel2012}  Seikel, M., Clarkson, C., \& Smith, M. 2012, JCAP, 06, 036
\bibitem[Shu et al.(2017)]{Shu17} Shu, Y., et al. 2017, arXiv:1711.00072
\bibitem[Sonnenfeld et al.(2012)]{Sonnenfeld12} Sonnenfeld, A., et al. 2012, ApJ, 752, 163
\bibitem[Sonnenfeld et al.(2013)]{Sonnenfeld13a} Sonnenfeld, A., Treu, T., Gavazzi, R., et al. 2013a, ApJ, 777, 98
\bibitem[Spergel et al.(2015)]{Spergel15} Spergel, D., et al., 2015, arXiv:1503.03757
\bibitem[Sullivan et al.(2000)]{Sullivan00} Sullivan, M., Ellis, R., Nugent, P., Smail, I., \& Madau, P. 2000, MNRAS, 319, 549
\bibitem[Suyu et al.(2010)]{Suyu10} Suyu, S. H., et al. 2010, ApJ, 711, 201
\bibitem[Suyu et al.(2012)]{Suyu12b} Suyu, S. H., et al. 2012, ApJ, 750, 10
\bibitem[Suyu et al.(2013)]{Suyu13} Suyu, S. H., et al. 2013, ApJ, 766, 70
\bibitem[Suyu et al.(2014)]{Suyu14} Suyu, S. H., et al. 2014, ApJ, 788, L35
\bibitem[Suyu et al.(2016)]{HOLI} Suyu, S. H., et al., arXiv:1607.00017
\bibitem[Suzuki et al.(2012)]{Suzuki12} Suzuki, N., et al. 2012, ApJ, 746, 85
\bibitem[Tewes et al.(2013a)]{Tewes13a} Tewes, M., Courbin, F., Meylan, G. 2013a, A\&A, 553, A120
\bibitem[Tewes et al.(2013b)]{Tewes13b} Tewes, M., et al. 2013b, A\&A, 556, A22
\bibitem[Treu et al.(2006)]{Treu06} Treu, T., et al. 2006b, ApJ, 650, 1219
\bibitem[Treu et al.(2010)]{Treu10} Treu, T. et al. 2010, ARA\&A, 48, 87
\bibitem[Walsh et al.(1979)]{Walsh79} Walsh, D., Carswell, R. F., \& Weymann, R. J. 1979, Nature, 279, 38
\bibitem[Young et al.(1981)]{Young81} Young, P., Gunn, J. E., Oke, J. B.,Westphal, J. A., \& Kristian, J. 1981, ApJ, 244, 736










\end{thebibliography}
\end{document}